\documentclass[aps, twocolumn, showpacs, 10pt, superscriptaddress, preprintnumbers, nofootinbib]{revtex4-1}
\usepackage[colorlinks, citecolor=blue]{hyperref}
\usepackage{enumerate}
\usepackage{xcolor, graphicx}
\usepackage{latexsym, cancel, amssymb, amsmath, verbatim, mathrsfs}
\usepackage{mathtools}
\usepackage{physics, ulem}
\usepackage{subfigure}
\usepackage{tikz-feynhand}

\usepackage{tikz-feynman} 

\newcommand{\beq}{\begin{equation}}
\newcommand{\eeq}{\end{equation}}
\newcommand{\nn}{\nonumber}

\newcommand{\bea}{\begin{eqnarray}}
\newcommand{\eea}{\end{eqnarray}}

\PassOptionsToPackage{normalem}{ulem}
\usepackage{ulem}
\RequirePackage{color}\definecolor{RED}{rgb}{1,0,0}\definecolor{BLUE}{rgb}{0,0,1} 

\allowdisplaybreaks[4]

\begin{document}

\preprint{
	{\vbox {
		\hbox{\bf CPTNP-2026-021}
}}}
\vspace*{0.2cm}

\title{Probing the CP violation effects via the angular coefficients in Drell-Yan production}

\author{Qingyi Situ}
\email{situqy@ihep.ac.cn}
\affiliation{Institute of High Energy Physics, Chinese Academy of Sciences, Beijing 100049, China
}
\affiliation{School of Physical Sciences, University
of Chinese Academy of Sciences, Beijing 100049, China}

\author{Bin Yan}
\email{yanbin@ihep.ac.cn (corresponding author)}
\affiliation{Institute of High Energy Physics, Chinese Academy of Sciences, Beijing 100049, China
}
\affiliation{Center for High Energy Physics, Peking University, Beijing 100871, China}

\begin{abstract}
The naive-$T$-odd angular coefficients $A_5$--$A_7$ in Drell--Yan production are highly suppressed in the Standard Model (SM), making them sensitive probes of CP-violating interactions beyond the SM. We study these observables within the Standard Model Effective Field Theory (SMEFT), focusing on dimension-six electroweak and chromomagnetic quark dipole operators. Due to their chirality-flipping structure, the leading contributions from CP-violating dipole interactions arise at $\mathcal{O}(1/\Lambda^4)$ through the interference between electroweak and chromomagnetic dipole amplitudes. We find that the sensitivity is primarily driven by $A_6$ and $A_7$ in the high dilepton transverse momentum region, where the dipole contributions are enhanced, while $A_5$ remains largely insensitive. Combining the ATLAS and CMS measurements, we constrain the CP-violating dipole combinations of Wilson coefficients at the $\mathcal{O}(0.1)$ level for $\Lambda=1~{\rm TeV}$. Assuming statistically dominated uncertainties at the HL-LHC, the sensitivity can be improved to the $\mathcal{O}(10^{-3})$ level. Our results highlight $A_6$ and $A_7$ as complementary probes of CP-violating dipole interactions at hadron colliders.
\end{abstract}

\maketitle

\section{Introduction}
The Drell–Yan production of lepton pairs at the Large Hadron Collider (LHC), $pp\to \gamma^*/Z\to \ell^+\ell^-+X$, is among the most precisely measured and theoretically well-controlled processes~\cite{Drell:1970wh}. Owing to its large production rate and clean experimental signature, it serves as a standard candle for precision tests of the Standard Model (SM) and provides a sensitive probe of possible new physics (NP) effects~\cite{Alioli:2018ljm,Alioli:2020kez,Boughezal:2021tih,Dawson:2021ofa,Boughezal:2022nof,Li:2022rag,Boughezal:2023nhe,Li:2024iyj,Grossi:2024tou,Grossi:2025aqj}. The angular distribution of the dilepton system offers direct access to the polarization of the intermediate gauge boson and encodes detailed information about the underlying electroweak interactions between gauge bosons and fermions. Within the SM, the angular distribution is parameterized by a set of frame-dependent angular coefficients $A_i$ ($i=0,\ldots,7$), which depend on the invariant mass, transverse momentum, and rapidity of the lepton pair and reflect the spin-1 nature of the exchanged gauge boson. Precise measurements of these coefficients therefore provide stringent tests of the SM while offering excellent sensitivity to NP contributions. A well-known example is the Lam--Tung relation, $A_0-A_2=0$~\cite{Lam:1978pu,Lam:1978zr,Lam:1980uc}, which holds at leading order and receives only small perturbative QCD corrections~\cite{Peng:2015spa}. However, significant violations of this relation have been observed by the ATLAS and CMS collaborations at $\sqrt{s}=8~\mathrm{TeV}$, and by the CMS and LHCb collaborations at $\sqrt{s}=13~\mathrm{TeV}$~\cite{ATLAS:2016rnf,LHCb:2022tbc,CMS:2026amb}, with noticeable deviations from the SM predictions at $\mathcal{O}(\alpha_s^3)$, particularly in the large transverse-momentum region~\cite{Gauld:2017tww}. These observations have motivated further investigations into whether the observed deviations arise from higher-order QCD and electroweak effects~\cite{Zhou:2009rp,Nefedov:2020ugj,Balitsky:2021fer,Frederix:2020nyw} or from NP contributions~\cite{Li:2024iyj,Grossi:2025aqj}, such as chirality-violating operators in the SMEFT, where heavy NP effects are systematically encoded in higher-dimensional operators constructed from SM fields while respecting the SM gauge symmetries~\cite{Buchmuller:1985jz,Grzadkowski:2010es}. More generally, physics beyond the SM can induce additional angular structures in the Drell--Yan process. For example, spherical harmonics with $l=3$ can arise through the interference between the SM amplitudes and dimension-eight SMEFT operators~\cite{Alioli:2020kez}.

Among the angular coefficients, $A_5$--$A_7$ are particularly interesting because they are odd under the naive-$T$ transformation, which reverses all momenta and spins without interchanging initial and final states. In CP-conserving theories, nonvanishing naive-$T$-odd observables originate solely from the absorptive part of the scattering amplitudes~\cite{osti_6368269}. Consequently, for the Drell--Yan process at nonzero transverse momentum, the leading SM contributions to $A_5$--$A_7$ arise only at $\mathcal{O}(\alpha_s^2)$ in perturbative QCD, rendering these observables highly suppressed. In contrast, sizeable effects can already appear at tree level if NP introduces new sources of CP violation~\cite{Petriello:2025lur}, making the naive-$T$-odd angular coefficients exceptionally clean probes of such interactions. 
The search for new sources of CP violation is strongly motivated by the observed baryon asymmetry of the universe, since it constitutes one of the three Sakharov conditions required for baryogenesis~\cite{Sakharov:1967dj}. Although CP violation has been firmly established in weak interactions through the Cabibbo--Kobayashi--Maskawa (CKM) matrix elements, its magnitude within the SM is insufficient to explain the observed baryon asymmetry. This strongly suggests the existence of additional sources of CP violation beyond the SM. Despite the availability of measurements of $A_5$--$A_7$ at the LHC, their potential as probes of CP-violating interactions has received relatively little attention.

In this work, we investigate CP-violating effects in the Drell--Yan process through the angular coefficients $A_5$--$A_7$, which have already been measured by the ATLAS and CMS collaborations at $\sqrt{s}=8~\mathrm{TeV}$~\cite{ATLAS:2016rnf} and $\sqrt{s}=13~{\rm TeV}$~\cite{CMS:2026amb}, respectively. Working within the SMEFT framework, we focus on the leading NP contributions arising from dimension-six operators of the form $C_i\mathcal{O}_i/\Lambda^2$, where the dimensionless Wilson coefficients $C_i$ parameterize the strengths of the corresponding interactions~\cite{Buchmuller:1985jz,Grzadkowski:2010es}. In particular, we consider the weak and chromomagnetic dipole operators, whose complex Wilson coefficients provide new sources of CP-violating phases. These dipole operators have previously been shown to play an important role in explaining the observed violation of the Lam--Tung relation~\cite{Li:2024iyj,Grossi:2025aqj}, and several complementary observables based on the transverse spin information of quarks and leptons have recently been proposed to probe them with enhanced sensitivity~\cite{Wen:2023xxc,Wen:2024cfu,Wen:2024nff,Huang:2025ljp,Cao:2025qua,Li:2025fom}. We demonstrate that nonvanishing contributions to the naive-$T$-odd angular coefficients $A_5$--$A_7$ arise only through the simultaneous presence of both weak and color dipole operators, providing a distinctive signature of their interference. Using the existing ATLAS and CMS measurements, we derive constraints on the relevant combinations of the real and imaginary parts of the corresponding Wilson coefficients, reaching the $\mathcal{O}(0.1)$ level for a NP scale of $\Lambda=1~\mathrm{TeV}$. We further show that the sensitivity could be improved to the $\mathcal{O}(10^{-3})$ level at the High-Luminosity LHC (HL-LHC) at $\sqrt{s}=14~{\rm TeV}$, demonstrating that the naive-$T$-odd angular coefficients constitute powerful and complementary probes of CP-violating interactions in the SMEFT.

\section{Angular coefficients in Drell-Yan process}
\begin{figure}[bt]
    \centering
    \includegraphics[width=.7\linewidth]{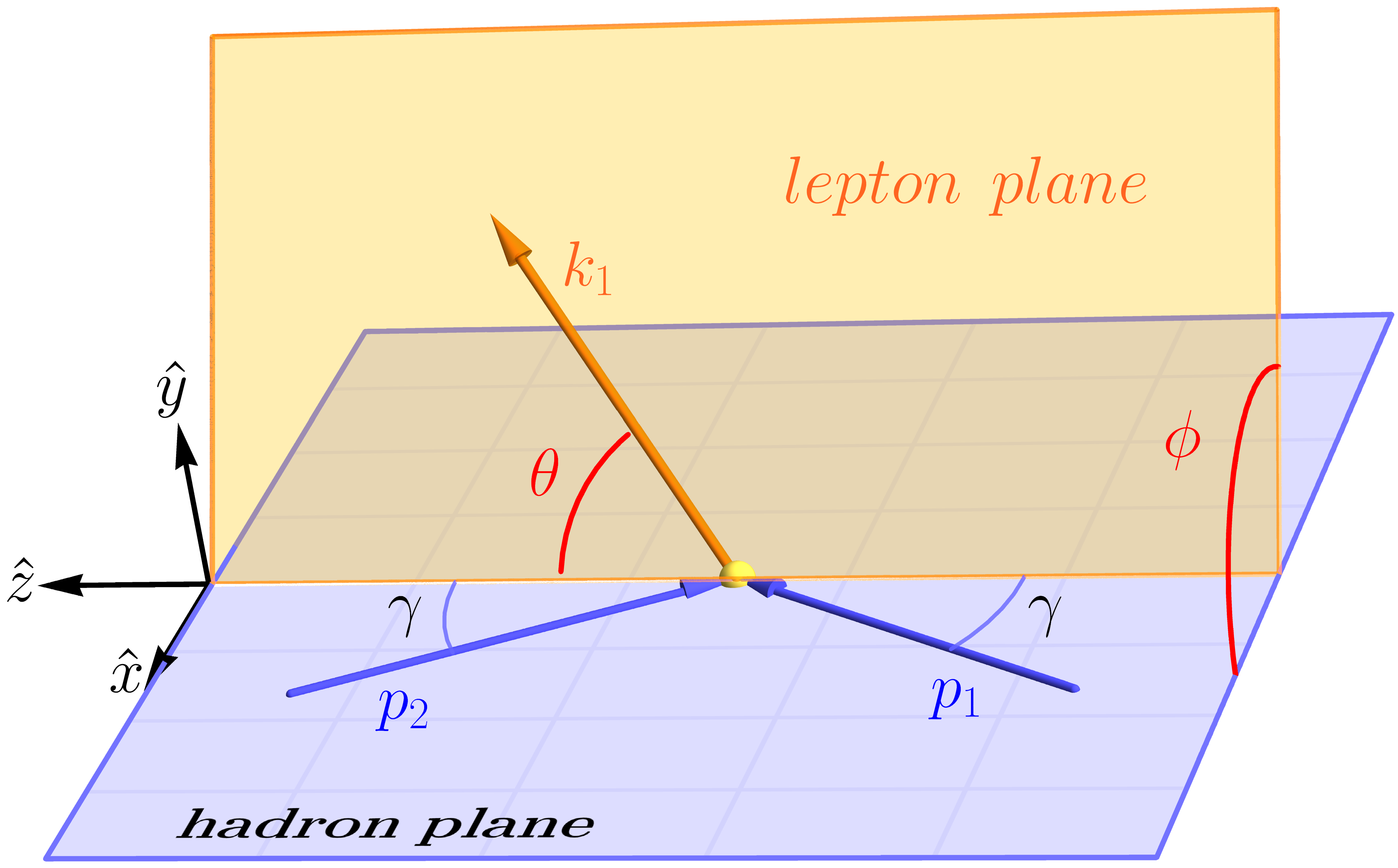}
    \caption{The definition of the Collins-Soper frame in the dilepton rest  frame and angles $\theta$ and $\phi$ are corresponding to the negatively charged lepton produced in the $\gamma^*/Z$ decay.}
    \label{fig:cs}
\end{figure}
The angular distributions of leptons in the Drell-Yan process at LHC can be expanded in terms of harmonic polynomial with $l\leq 2$ in the Collins-Soper (CS) frame~\cite{Collins:1977iv},
\bea
&&\frac{d\sigma}{dp_T^{\ell\ell}dy_{\ell\ell}dm_{\ell\ell\ }^2d\cos{\theta}d\phi\ } \nonumber \\
&&=\frac{3}{16\pi}\frac{d\sigma}{dp_T^{\ell\ell}dy_{\ell\ell}dm_{\ell\ell}^2} \left\{\left(1+\cos^2{\theta}\right)+\frac{1}{2}A_0\left(1-3\cos^2{\theta}\right) \right. \nonumber \\ 
&& +A_1\sin{2\theta}\cos{\phi}+\frac{1}{2}A_2\sin^2{\theta}\cos{2\phi}+A_3\sin{\theta}\cos{\phi} \nonumber \\
&&  +A_4 \cos{\theta}  +A_5\sin^2{\theta}\sin{2\phi}+A_6\sin{2\theta}\sin{\phi}  \nonumber \\
&&  \left. +A_7\sin{\theta}\sin{\phi} \right\} \;,
\label{eq:expansion}
\eea
where $\theta$ and $\phi$ denote the polar and azimuthal angles of the negatively charged lepton in the CS frame, illustrated in Fig.~\ref{fig:cs}, while $y_{\ell\ell}$ and $m_{\ell\ell}$ are the rapidity and invariant mass of the dilepton system, respectively. Owing to the orthogonality of the harmonic polynomials, the angular coefficients can be projected out from the differential cross section. In particular,
\bea
&& A_5=5\left \langle \sin^2{\theta}\sin{2\phi} \right \rangle,\quad A_6=5\left \langle \sin{2\theta}\sin{\phi} \right \rangle,\nn\\
&&A_7=4\left \langle \sin{\theta}\sin{\phi} \right \rangle,
\eea
where
\begin{equation}
\left\langle
f(\theta,\phi)
\right\rangle
=
\frac{
\int d\Omega\,
d\sigma(\theta,\phi)\,
f(\theta,\phi)
}{
\int d\Omega\,
d\sigma(\theta,\phi)
}.
\end{equation}

The angular coefficients can also be expressed in terms of the normalized spin-density matrix of the intermediate gauge boson,
\begin{equation}
    \begin{aligned}
    &
    \quad\quad\rho_{\lambda\lambda^{'}}=
    \\&
        \begin{pmatrix}
            \frac{1}{3}+\frac{\delta_L}{6}+\frac{J_3}{2} & \frac{J_1+Q_{xz}-i({\color{black}J_2}+{\color{black}Q_{yz}})}{2\sqrt{2}} & \frac{\lambda_T-i{\color{black}Q_{xy}}}{2} \\
            \frac{J_1+Q_{xz}+i(J_2+Q_{yz})}{2\sqrt{2}} & \frac{1-\delta_L}{3} & \frac{J_1-Q_{xz}-i(J_2-Q_{yz})}{2\sqrt{2}} \\
            \frac{\lambda_T+iQ_{xy}}{2} & \frac{J_1-Q_{xz}+i(J_2-Q_{yz})}{2\sqrt{2}} & \frac{1}{3}+\frac{\delta_L}{6}-\frac{J_3}{2}
        \end{pmatrix},
    \end{aligned}
\end{equation}
where $J_i$ denote the polarization components of the gauge boson along the $i$-th direction and $Q_{ij}$ denote its quadrupole moments. The diagonal elements correspond to the helicity fractions of the gauge boson, while the off-diagonal elements encode the interference between different helicity states. Combining the production density matrix with the leptonic decay amplitudes yields
\begin{equation}
A_5\propto Q_{xy},\qquad
A_6\propto Q_{yz},\qquad
A_7\propto J_2.
\end{equation}
Consequently,  the coefficients $A_5$--$A_7$ probe the imaginary components of the spin-density matrix, which arise from absorptive phases in the SM or from new sources of CP violation beyond the SM.

In the SMEFT framework, deviation of $A_5$--$A_7$  from their SM predictions at $\mathcal{O}(\alpha_s^2)$ can therefore provide a probe of dimension-six operators with new CP-violating phases. Since these observables require imaginary parts of the production amplitudes, only operators that introduce complex phases into the production amplitudes can contribute at tree level. At dimension six, this requirement uniquely singles out the quark dipole operators,  whose complex Wilson coefficients provide additional sources of CP violation. Related analyses based on CP-odd dimension-eight operators have recently been explored in Ref.~\cite{Petriello:2025lur}.

Within the Warsaw basis of the SMEFT~\cite{Grzadkowski:2010es}, the relevant dipole operators are
\begin{equation}
\begin{aligned}
\mathcal{O}_{uG}&=
(\bar q_p\sigma^{\mu\nu}T^Au_r)
\tilde{\varphi}
G_{\mu\nu}^A,
&
\mathcal{O}_{dG}&=
(\bar q_p\sigma^{\mu\nu}T^Ad_r)
\varphi
G_{\mu\nu}^A,
\\
\mathcal{O}_{uW}&=
(\bar q_p\sigma^{\mu\nu}u_r)
\tau^I
\tilde{\varphi}
W_{\mu\nu}^I,
&
\mathcal{O}_{dW}&=
(\bar q_p\sigma^{\mu\nu}d_r)
\tau^I
\varphi
W_{\mu\nu}^I,
\\
\mathcal{O}_{uB}&=
(\bar q_p\sigma^{\mu\nu}u_r)
\tilde{\varphi}
B_{\mu\nu},
&
\mathcal{O}_{dB}&=
(\bar q_p\sigma^{\mu\nu}d_r)
\varphi
B_{\mu\nu},
\end{aligned}
\end{equation}
where $q_p$ denotes the left-handed quark doublet, $u_r$ and $d_r$ denote the right-handed up- and down-type quark singlets, $\varphi$ is the Higgs doublet, and $G_{\mu\nu}$, $W_{\mu\nu}$, and $B_{\mu\nu}$ are the field-strength tensors associated with the $SU(3)_C$, $SU(2)_L$, and $U(1)_Y$ gauge groups, respectively.

After electroweak symmetry breaking, the neutral electroweak dipole operators can be expressed in terms of the physical $Z$ boson and photon fields,
\begin{equation}
\begin{aligned}
\mathcal{O}_{uZ}&=
\frac{v}{\sqrt2}
(\bar u_p\sigma^{\mu\nu}u_r)
Z_{\mu\nu},
&
\mathcal{O}_{dZ}&=
\frac{v}{\sqrt2}
(\bar d_p\sigma^{\mu\nu}d_r)
Z_{\mu\nu},
\\
\mathcal{O}_{uA}&=
\frac{v}{\sqrt2}
(\bar u_p\sigma^{\mu\nu}u_r)
A_{\mu\nu},
&
\mathcal{O}_{dA}&=
\frac{v}{\sqrt2}
(\bar d_p\sigma^{\mu\nu}d_r)
A_{\mu\nu},
\end{aligned}
\end{equation}
where
\begin{equation}
C_{uZ}
=
-s_WC_{uB}
+c_WC_{uW},
\quad
C_{dZ}
=
-s_WC_{dB}
-c_WC_{dW}.
\end{equation}
Since our analysis focuses on the $Z$-pole region, $m_{\ell\ell}\in [80,100]~{\rm GeV}$, the contributions from the photon dipole operators are suppressed and will therefore be neglected throughout this work.

The chirality-flipping nature of dipole operators suppresses their interference with the SM amplitudes. Consequently, neither the electroweak nor the chromomagnetic dipole operator alone can generate nonvanishing contributions to $A_5$--$A_7$ at tree level. The leading effects arise from the mixed interference between electroweak and chromomagnetic dipole operators at $\mathcal O(1/\Lambda^4)$. A nonzero contribution requires two chirality flips along the fermion line, such that the two dipole insertions correspond to one operator and the Hermitian conjugate of the other. These insertions can occur either on opposite sides of the cut diagram, corresponding to amplitude and conjugate amplitude interference, or within the same amplitude. Consequently, the relevant CP-violating combinations of Wilson coefficients contributing to the $A_5$--$A_7$ are given by  $C_u\equiv {\rm Im}(C_{uG}C_{uZ}^*)$ and  $C_d\equiv {\rm Im}(C_{dG}C_{dZ}^*)$.

\section{Numerical results and discussion}

\begin{figure}[bt]
    \centering
    \includegraphics[width=0.9\linewidth]{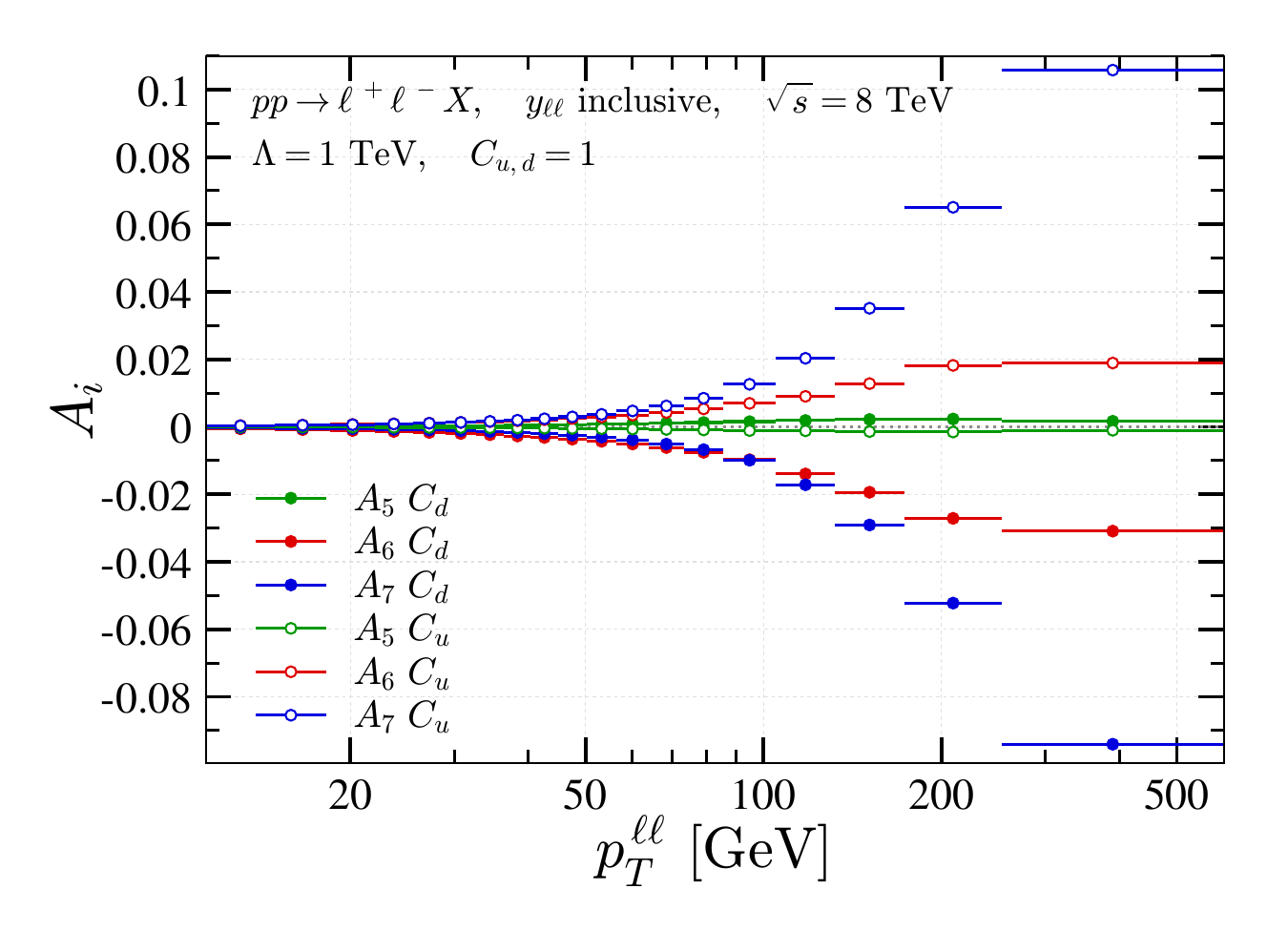}
    \caption{The contribution of dipole operators to the angular coefficients $A_5$--$A_7$ as a function of $p_T^{\ell\ell}$ at $\sqrt{s}=8~{\rm TeV}$ LHC, with $C_{u,d}=1$ and $\Lambda=1~{\rm TeV}$.}
    \label{fig:Ai8}
\end{figure}

\begin{figure*}[bt]
    \centering
    \includegraphics[width=0.45\linewidth]{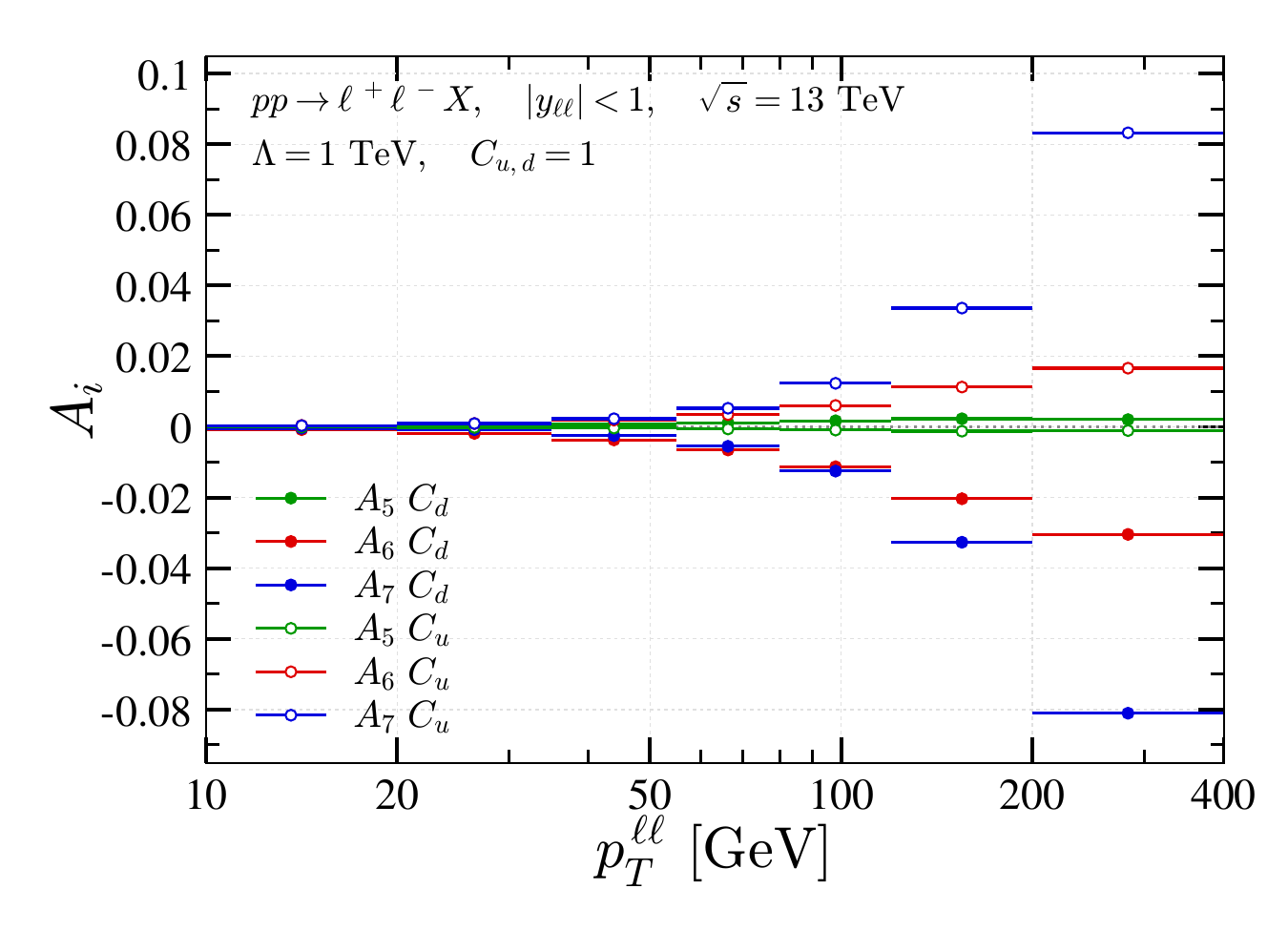}
    \includegraphics[width=0.45\linewidth]{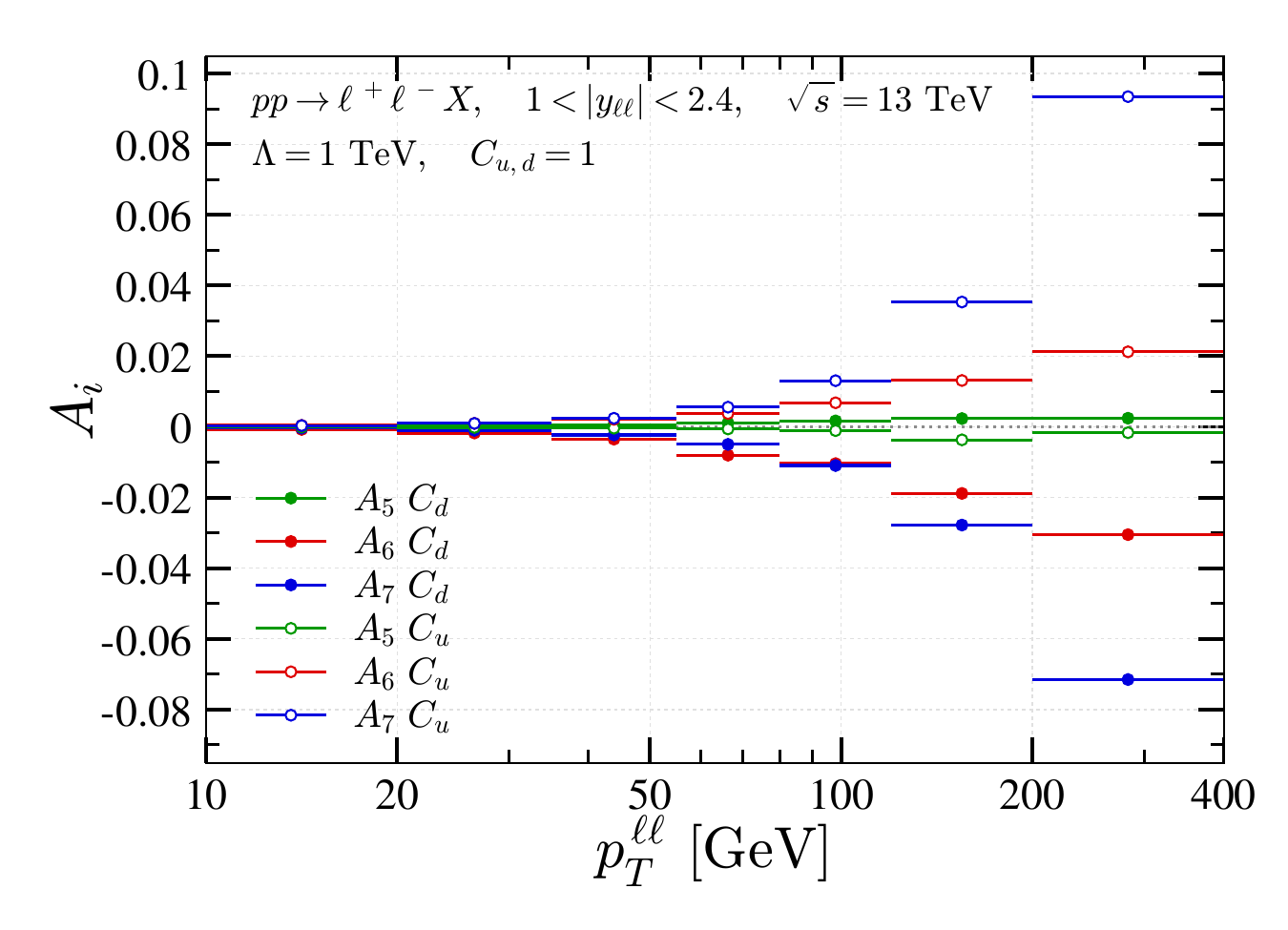}
    \caption{Same as Fig.~\ref{fig:Ai8}, but for $\sqrt{s}=13~{\rm TeV}$ with different rapidity of the dilepton.}
    \label{fig:Ai13}
\end{figure*}

We now present the projected sensitivity to the dipole operators from measurements of the angular coefficients $A_5$--$A_7$ in Drell--Yan production at the LHC. We adopt the transverse mass of the dilepton system~\cite{Gauld:2017tww},
\begin{equation}
m_T^Z=\sqrt{m_{\ell\ell}^2+\left(p_T^{\ell\ell}\right)^2},
\end{equation}
as the central renormalization and factorization scale. This dynamical scale choice is particularly suitable for the high-$p_T^{\ell\ell}$ region considered here, where the relevant hard scale of the process is set by both the dilepton invariant mass and its transverse momentum, rather than by $m_{\ell\ell}$ alone. Figure~\ref{fig:Ai8} shows the dipole contributions to $A_5$--$A_7$ at $\sqrt{s}=8~{\rm TeV}$ for $C_{u,d}=1$ and $\Lambda=1~{\rm TeV}$ in the invariant-mass window $m_{\ell\ell}\in[80,100]~{\rm GeV}$. The dipole contribution to $A_5$ is found to be negligible compared with those to $A_6$ and $A_7$, consistent with Ref.~\cite{Petriello:2025lur}. 
In addition, the up- and down-type dipole operators contribute with opposite signs due to the different $Z$-boson couplings to up- and down-type quarks. At large $p_T^{\ell\ell}$, they induce sizable effects in $A_6$ and $A_7$, which remain largely insensitive to the collider energy and dilepton rapidity selection, as illustrated by the $\sqrt{s}=13~{\rm TeV}$ predictions in Fig.~\ref{fig:Ai13}.

We next perform a $\chi^2$ analysis to constrain the Wilson coefficients of the dipole operators using the measured angular coefficients $A_5$--$A_7$ at the LHC. We include the ATLAS measurements at $\sqrt{s}=8~{\rm TeV}$ with an integrated luminosity of $20.3~{\rm fb}^{-1}$ for both $Z\to e^+e^-$ and $Z\to\mu^+\mu^-$ final states~\cite{ATLAS:2016rnf}, as well as the corresponding CMS measurements at $\sqrt{s}=13~{\rm TeV}$ with an integrated luminosity of $140~{\rm fb}^{-1}$ in the $Z\to\mu^+\mu^-$ channel~\cite{CMS:2026amb}. We define
\begin{equation}
\chi^2
=
\sum_{i=5}^{7}\sum_b
\frac{
\left(
A_{i,b}^{\rm exp}
-
A_{i,b}^{\rm SM}
-
A_{i,b}^{\rm SMEFT}
\right)^2
}{
\left(\delta A_{i,b}^{\rm exp}\right)^2
+
\left(\delta A_{i,b}^{\rm th}\right)^2
},
\tag{10}
\end{equation}
where $b$ labels the $p_T^{\ell\ell}$ and $y_{\ell\ell}$ bins included in the ATLAS and CMS measurements. Here, $A_{i,b}^{\rm exp}$ denotes the measured angular coefficient, $A_{i,b}^{\rm SM}$ is the SM prediction evaluated at $\mathcal{O}(\alpha_s^2)$, and $A_{i,b}^{\rm SMEFT}$ denotes the contribution from the dipole operators, calculated at tree level and retained at $\mathcal{O}(1/\Lambda^4)$ in the SMEFT expansion. The experimental uncertainty $\delta A_{i,b}^{\rm exp}$ includes both statistical and systematic components, while $\delta A_{i,b}^{\rm th}$ accounts for the theoretical uncertainty associated with the QCD scale variation.

To derive the individual bounds for the dipole operators, we consider two scenarios in this work:
\begin{itemize}
\item $A_i^{\rm exp}$ corresponds to the current ATLAS measurements with regularized data at $\sqrt{s}=8~{\rm TeV}$ and CMS measurements at $\sqrt{s}=13~{\rm TeV}$, incorporating both the theoretical and experimental uncertainties in $\chi^2$ analysis;
\item Pseudo-experiments for the HL-LHC are generated using the SM predictions at $\mathcal{O}(\alpha_s^2)$ as the central values. We consider both $Z\to e^+e^-$ and $Z\to \mu^+\mu^-$ channels, assume purely statistical uncertainties, and rescale the corresponding uncertainties according to the luminosity-scaling prescription in Ref.~\cite{ATLAS:2019mfr}.
\end{itemize}

\begin{figure}[bt]
    \centering
    \includegraphics[width=1.0\linewidth]{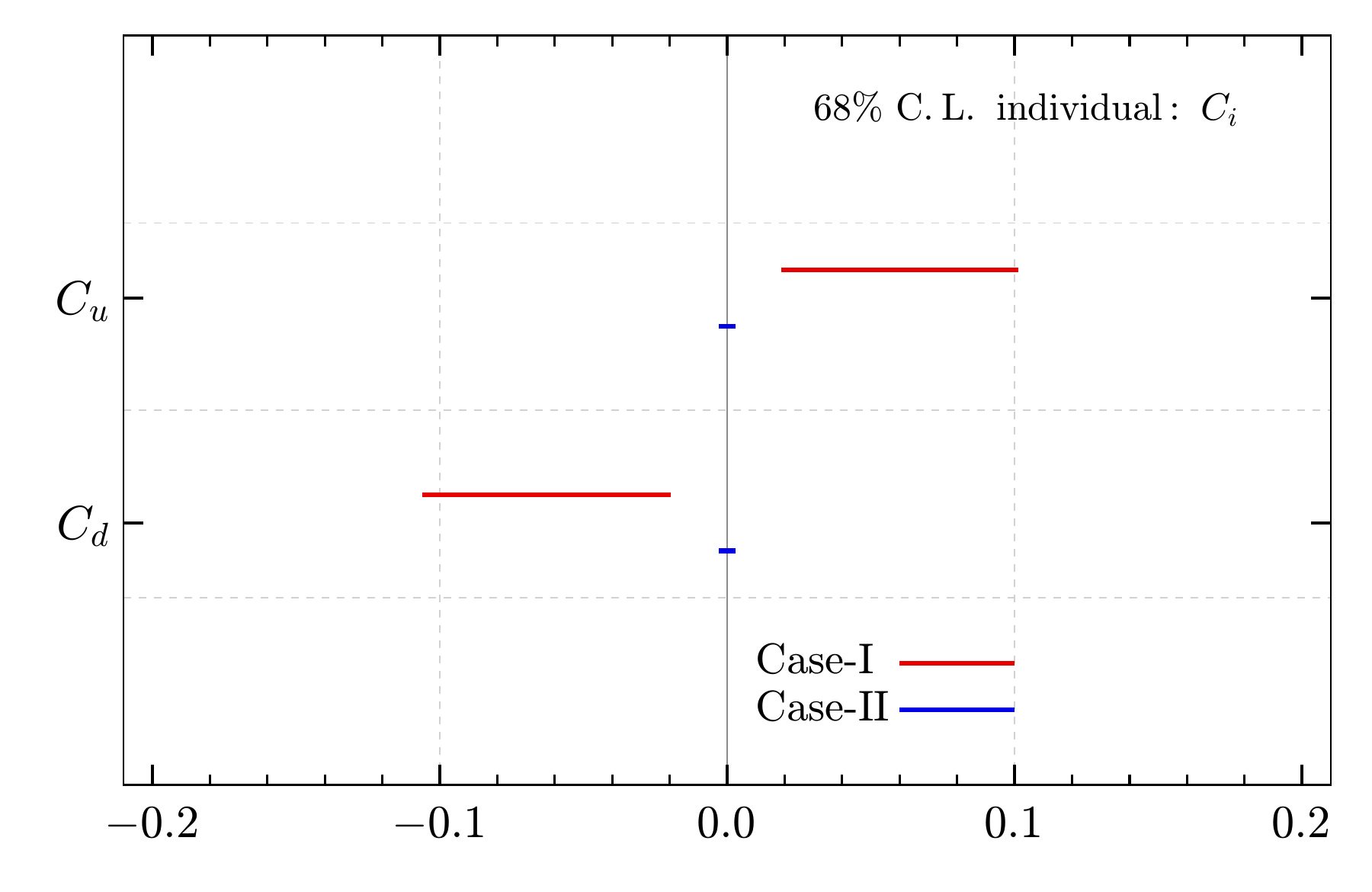}
    \caption{The individual bounds for the Wilson coefficients of dipole operators at $68\%$  C.L. with $\Lambda=1~{\rm TeV}$.}
    \label{fig:result}
\end{figure}

For case-I, the red lines in Fig.~\ref{fig:result} show the 68\% C.L. constraints on the Wilson coefficients when each dipole coefficient is varied individually, with the others set to zero. The opposite signs of the up- and down-type quark constraints originate from the different $Z$-boson couplings, which lead to opposite contributions to $A_5$--$A_7$, as shown in Figs.~\ref{fig:Ai8} and~\ref{fig:Ai13}. For $\Lambda=1~{\rm TeV}$, the relevant combinations of Wilson coefficients are constrained at the $\mathcal{O}(0.1)$ level.

\begin{figure*}[bt]
    \centering
    \includegraphics[width=0.45\linewidth]{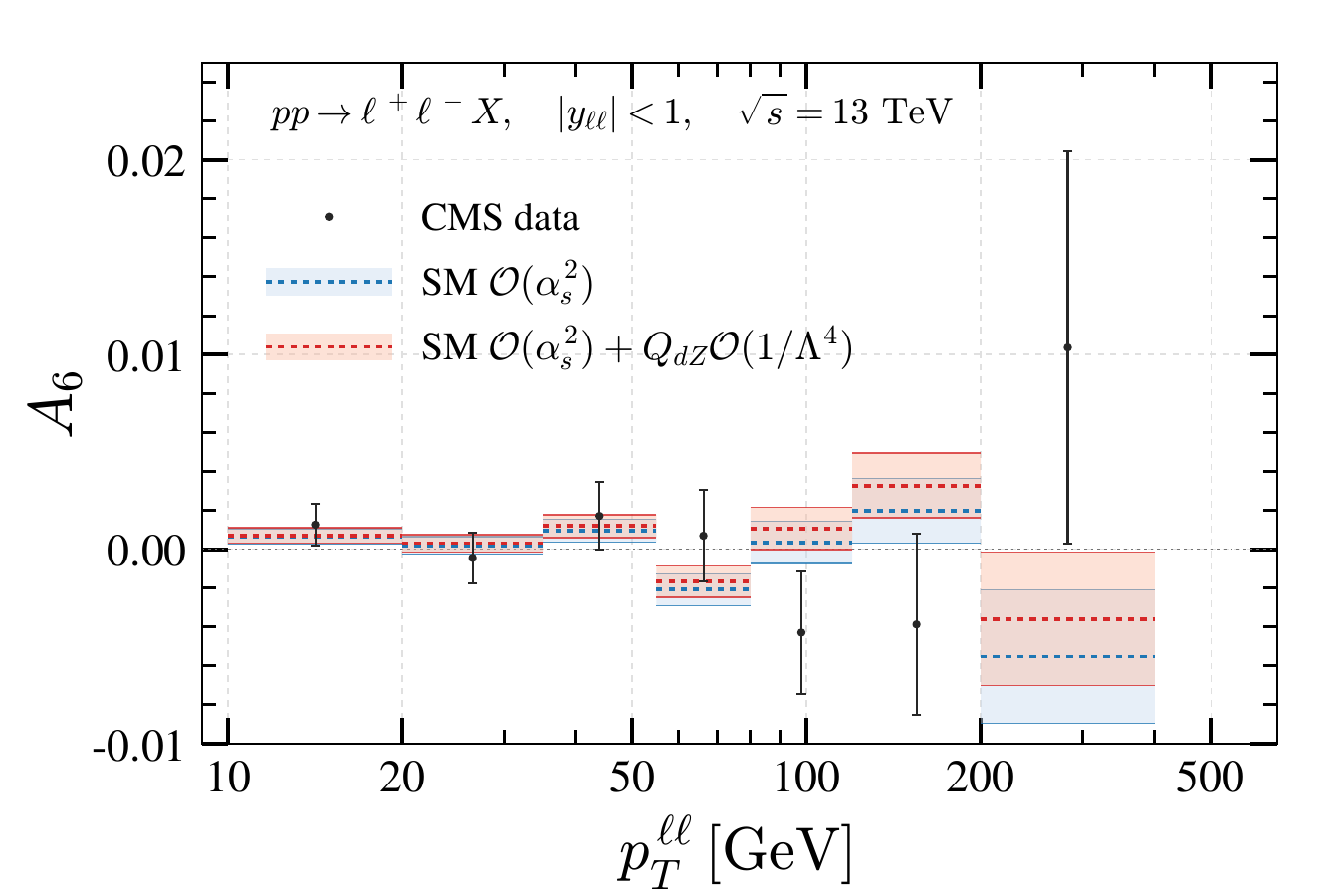}
    \includegraphics[width=0.45\linewidth]{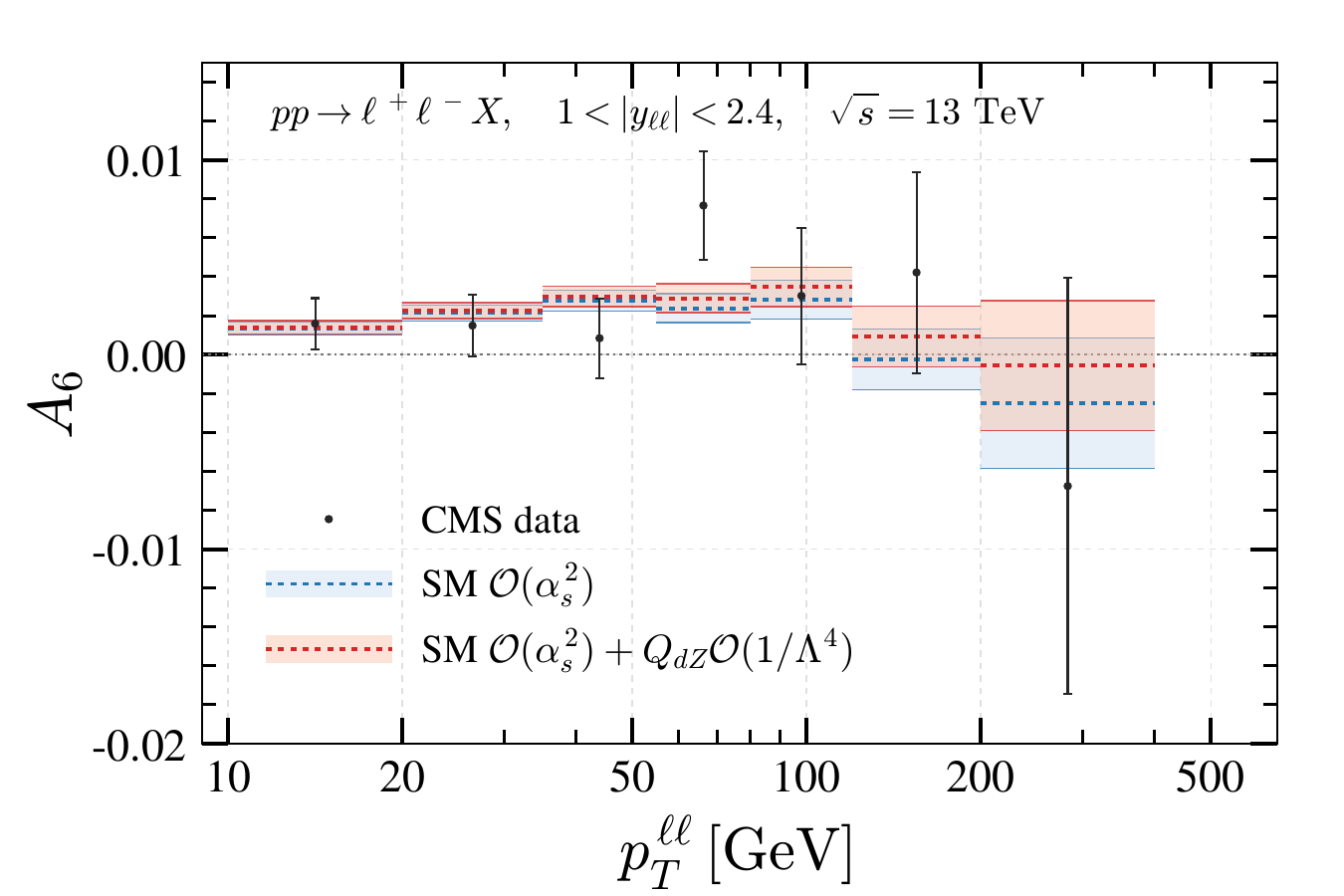}
     \includegraphics[width=0.45\linewidth]{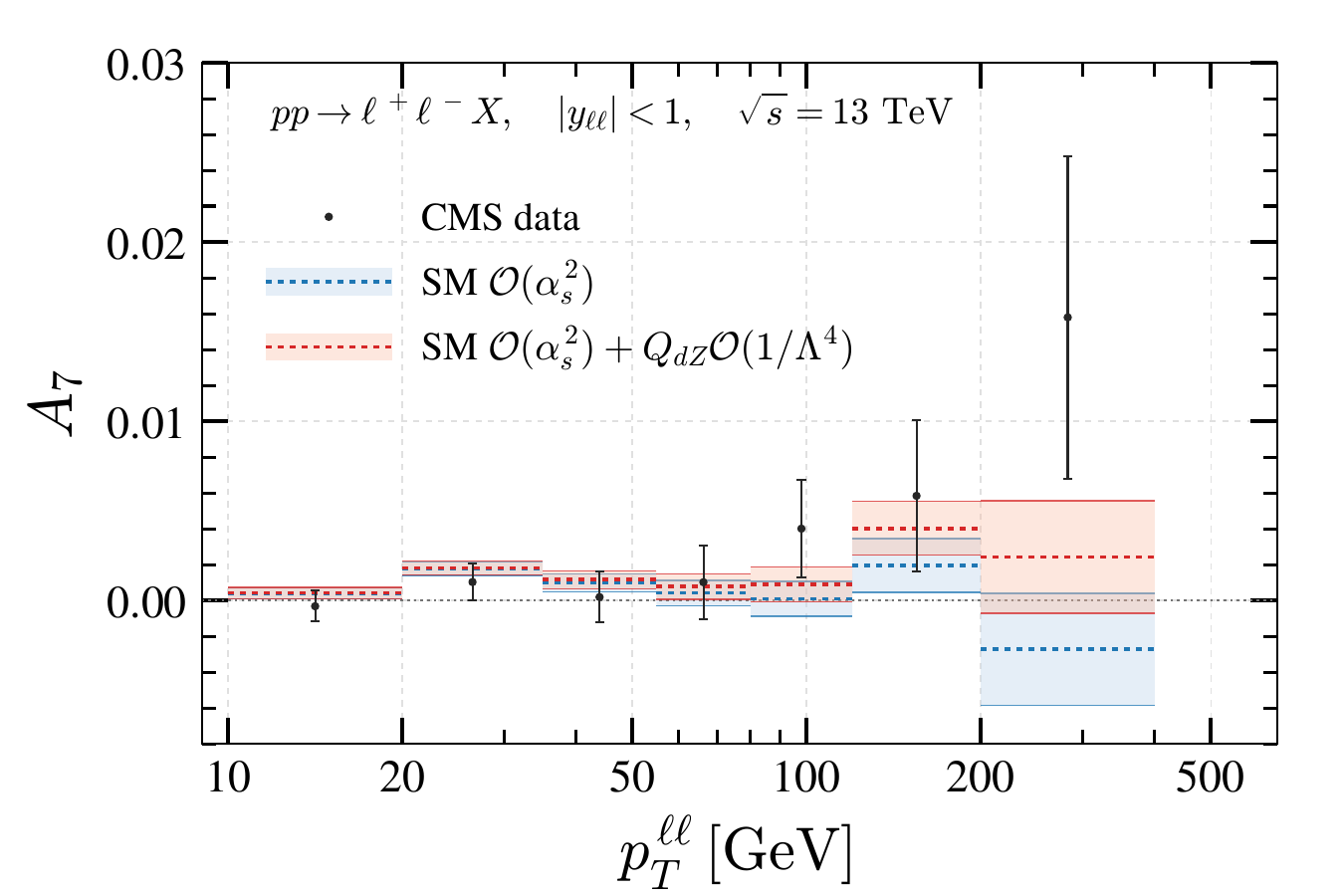}
    \includegraphics[width=0.45\linewidth]{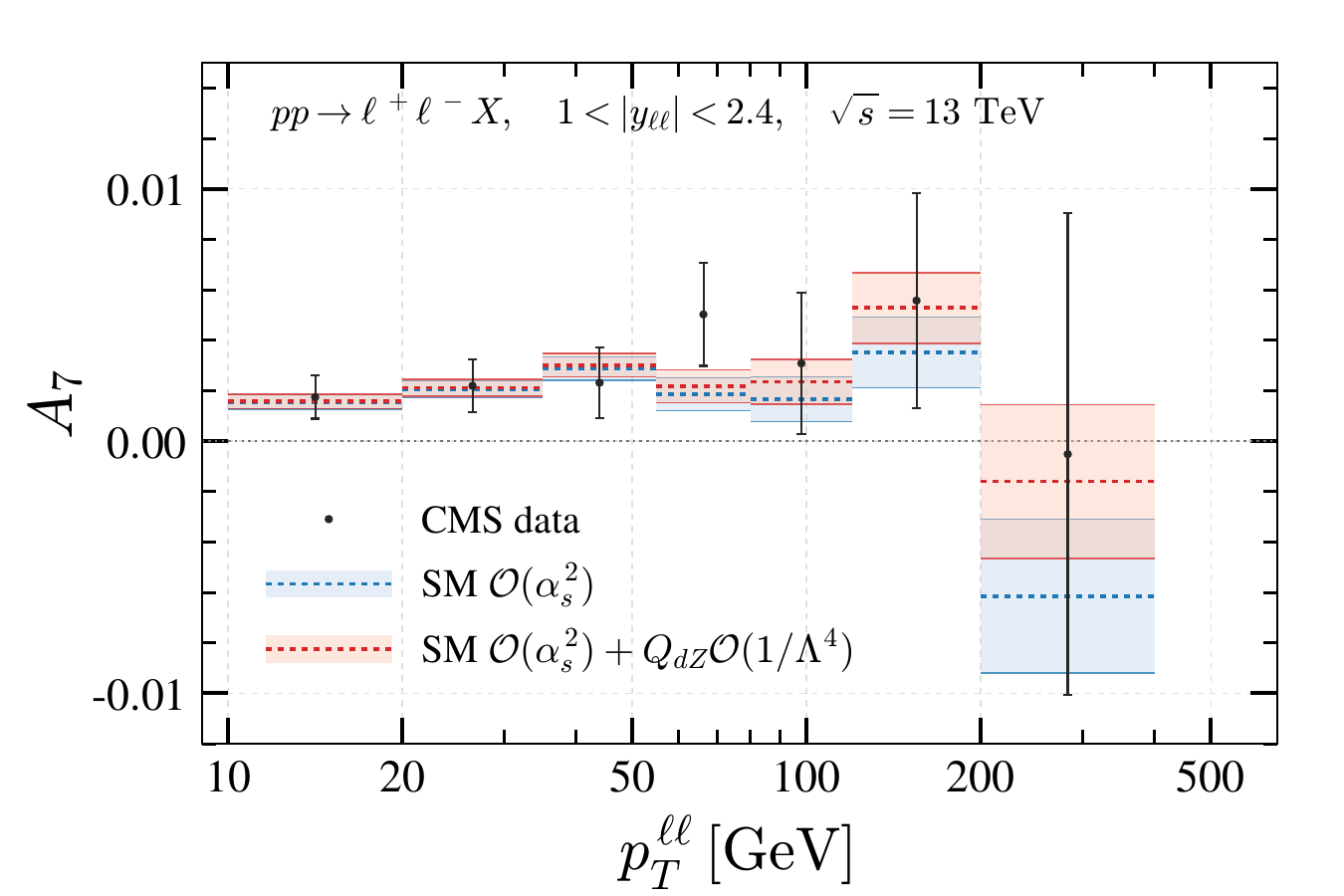}
    \caption{The $p_T^{\ell\ell}$ distributions of the angular coefficients $A_6$ and $A_7$ at the $\sqrt{s}=13~{\rm TeV}$ LHC. The CMS measurements~\cite{CMS:2026amb} (black points) are compared with the SM predictions at $\mathcal{O}(\alpha_s^2)$ accuracy~\cite{CMS:2026amb} (blue bands) and the combined SM prediction including the down-type dipole contribution with the best-fit Wilson coefficient obtained from Fig.~\ref{fig:result} (red bands).}
    \label{fig:Ai13d}
\end{figure*}

In Fig.~\ref{fig:Ai13d}, we compare the $p_T^{\ell\ell}$ dependence of $A_6$ and $A_7$ among the CMS measurements (black points), the SM predictions at $\mathcal{O}(\alpha_s^2)$ accuracy (blue bands), and the predictions including the down-type dipole contribution with the best-fit value of $C_d$ (red bands) at $\sqrt{s}=13~{\rm TeV}$. We focus on the CMS measurements since they provide smaller uncertainties compared with the corresponding ATLAS measurements at $\sqrt{s}=8~{\rm TeV}$. It is evident that including the dipole contribution improves the agreement between the theoretical predictions and the measured values of $A_6$ and $A_7$, particularly in the high-$p_T^{\ell\ell}$ region. The corresponding results for the up-type dipole operator exhibit similar behavior and are not shown.

For case-II, we evaluate the SM cross section at $\sqrt{s}=14~{\rm TeV}$ at $\mathcal{O}(\alpha_s)$ accuracy. The higher-order QCD effects are estimated by applying the ratio $\kappa=\sigma(\mathcal O(\alpha_s^3))/\sigma(\mathcal O(\alpha_s))$
obtained at $\sqrt{s}=13~{\rm TeV}$ in Ref.~\cite{Boughezal:2015ded} to the HL-LHC prediction. This approximation is supported by the weak energy dependence of $\kappa$ between  $\sqrt{s}=8$ and $13~{\rm TeV}$~\cite{Boughezal:2015ded}. The projected HL-LHC constraints could probe the dipole operator combinations at the $\mathcal{O}(10^{-3})$ level, improving the current sensitivity by about two orders of magnitude.

The CP-violating combinations of quark dipole operators are also constrained by low-energy EDM measurements, particularly through their contributions to neutron and atomic EDMs after renormalization-group evolution and matching onto hadronic operators~\cite{Pitschmann:2014jxa,Liu:2017olr,Mereghetti:2018oxv,Kley:2021yhn,ParticleDataGroup:2026aaa}. However, the resulting bounds depend on theoretical assumptions and suffer from uncertainties associated with nonperturbative hadronic matrix elements. In contrast, the angular coefficients $A_5$--$A_7$ directly probe the interference of electroweak and chromomagnetic dipole interactions at high energies. The Wilson coefficient combinations probed by these observables are not identical to those constrained by low-energy EDMs, making collider measurements complementary to low-energy searches.

\section{Conclusions}
In this work, we investigate the sensitivity of the naive-$T$-odd angular coefficients $A_5$--$A_7$ in Drell--Yan production to CP-violating quark dipole interactions within the SMEFT framework. Since the SM contributions to these observables require absorptive phases, they first arise at $\mathcal{O}(\alpha_s^2)$, making $A_5$--$A_7$ clean probes of physics beyond the SM. We show that the leading tree-level SMEFT contributions from dimension-six quark dipole operators arise at $\mathcal{O}(1/\Lambda^4)$ through the interference between electroweak $Z$-dipole and chromomagnetic dipole operators. These contributions are governed by the CP-violating Wilson coefficient combinations $C_u={\rm Im}(C_{uG}C_{uZ}^*)$ and $C_d={\rm Im}(C_{dG}C_{dZ}^*)$. The sensitivity is dominated by $A_6$ and $A_7$ at large $p_T^{\ell\ell}$, while $A_5$ receives negligible contributions. Using existing ATLAS and CMS measurements, we constrain these Wilson coefficient combinations at the $\mathcal{O}(0.1)$ level for $\Lambda=1~{\rm TeV}$. Under the assumption of statistically dominated uncertainties, the HL-LHC projection can improve the sensitivity to the $\mathcal{O}(10^{-3})$ level. These results highlight high-$p_T^{\ell\ell}$ measurements of $A_6$ and $A_7$ as complementary probes of CP-violating quark dipole interactions at hadron colliders.

\vspace{3mm}
\noindent{\bf Acknowledgments.}
The authors were partly supported by the National Natural Science Foundation of China under Grant No.~12422506, and CAS under Grant No.~E429A6M1. The authors gratefully acknowledge the valuable discussions and insights provided by the members of the Collaboration on Precision Tests and New Physics (CPTNP).

\bibliographystyle{apsrev}
\bibliography{reference}

\end{document}